\documentclass[final,5p,times,twocolumn,authoryear]{elsarticle}

\usepackage{amsthm} \usepackage{amssymb}
\usepackage{lipsum}
\usepackage{amsmath}
\usepackage{color}
\usepackage{CJKutf8}
\usepackage{natbib}
\setcitestyle{numbers,square}
\journal{Physics Letters B}

\begin{document}
\begin{CJK}{UTF8}{gbsn}

\begin{frontmatter}

%% Title, authors and addresses

%% use the tnoteref command within \title for footnotes;
%% use the tnotetext command for theassociated footnote;
%% use the fnref command within \author or \affiliation for footnotes;
%% use the fntext command for theassociated footnote;
%% use the corref command within \author for corresponding author footnotes;
%% use the cortext command for theassociated footnote;
%% use the ead command for the email address,
%% and the form \ead[url] for the home page:
%% \title{Title\tnoteref{label1}}
%% \tnotetext[label1]{}
%% \author{Name\corref{cor1}\fnref{label2}}
%% \ead{email address}
%% \ead[url]{home page}
%% \fntext[label2]{}
%% \cortext[cor1]{}
%% \affiliation{organization={},
%%            addressline={}, 
%%            city={},
%%            postcode={}, 
%%            state={},
%%            country={}}
%% \fntext[label3]{}

\title{Bayesian method for quantifying the non-Gaussian fluctuations in low-intermediate energy heavy ion collisions}

%% use optional labels to link authors explicitly to addresses:
%% \author[label1,label2]{}
%% \affiliation[label1]{organization={},
%%             addressline={},
%%             city={},
%%             postcode={},
%%             state={},
%%             country={}}
%%
%% \affiliation[label2]{organization={},
%%             addressline={},
%%             city={},
%%             postcode={},
%%             state={},
%%             country={}}

% \author[first]{Author name}
% \affiliation[first]{organization={University of the Moon},%Department and Organization
%             addressline={}, 
%             city={Earth},
%             postcode={}, 
%             state={},
%             country={}}
\author[label1]{Xinyu Wang}
\author[label1]{Xiang Chen}
\author[label1]{Junping Yang}
\author[label1]{Ying Cui\corref{cor1}}
\ead{yingcuid@163.com}
\author[label1]{Zhuxia Li}
\author[label1]{Kai Zhao\corref{cor1}}
\ead{zhaokai@ciae.ac.cn}
\author[label1,label2,label3]{Yingxun Zhang\corref{cor1}}
\ead{zhyx@ciae.ac.cn}
\affiliation[label1]{China Institute of Atomic Energy, P. O. Box 275(18), Beijing 102413, China}
\affiliation[label2]{Department of Physics and Technology, Guangxi normal University, Guilin 540101, China}
\affiliation[label3]{Shanghai Research Center for Theoretical Nuclear Physics, NSFC and Fudan University, Shanghai 200438, China}

\cortext[cor1]{Corresponding author}

\begin{abstract}
%% Text of abstract
% Example abstract for the Physics Letters B journal. Here you provide a brief summary of the research and the results.
In this work, %the Bayesian method for reconstructing the impact parameter distributions from the measured nucleons and light particles is investigated under the conditions of experimental filters. Our results show that the Bayesian method can well reconstruct the impact parameter distributions for the selected centrality according to the measured nucleons and particles. Furthermore, 
we present a model-independent method to quantify the non-Gaussian fluctuations in the observable distributions, which are assessed by the difference between the measured observable distributions and reconstructed observable distributions via the Bayesian method. Our results indicate that the strength of non-Gaussian fluctuation increases with the beam energy, and is primarily driven by non-central collision mechanisms. The experimental measurement of the strength of non-Gaussian fluctuation of the observable distributions will provide valuable insights into understanding the nonequilibrium effects in heavy ion collisions and the liquid-gas phase transition in finite nuclei.
\end{abstract}

%%Graphical abstract
%\begin{graphicalabstract}
%\includegraphics{grabs}
%\end{graphicalabstract}

%%Research highlights
%\begin{highlights}
%\item Research highlight 1
%\item Research highlight 2
%\end{highlights}

\begin{keyword}
%% keywords here, in the form: keyword \sep keyword, up to a maximum of 6 keywords
Bayesian method \sep non-Gaussian fluctuations \sep impact parameter distributions \sep observable distributions  %\sep keyword 4

%% PACS codes here, in the form: \PACS code \sep code

%% MSC codes here, in the form: \MSC code \sep code
%% or \MSC[2008] code \sep code (2000 is the default)

\end{keyword}

\end{frontmatter}

The Bayesian method is a statistical approach that involves using probability to represent uncertainty in the knowledge about model parameters and is widely used in algorithms of statistical analysis and machine learning\citep{Hewanbing2023}. For example, the Bayesian inference on the nuclear equation of state\citep{Scott2015,Morfouace2019,WJXie2021,JunXu2020}, a Bayesian Neural Network for approximating the nuclear fission yield data and nuclear mass\citep{Utama2016,peijunc2019,NIU201848}, a Naive Bayes Classifier for predicting nuclear charge radii and mass\citep{MaYunfei2020,LiuYifan2021}, Richard-Lucy algorithm for deblurring and accurately identifying sources\citep{Danielewicz2022}. Recently, the Bayesian method for reconstructing the impact parameter distributions and understanding the fluctuation mechanism for intermediate and high energy heavy ion collisions (HICs) were performed\citep{Frankland21,Das18,Rogly18,Yousefnia22}. %In reconstructing the impact parameter distributions and understand the fluctuation mechanism, the Bayesian method accounts for the fluctuation of observables with respect to impact parameter and can reconstruct the impact parameter distributions from the measured experimental observables in a model-independent way. Yousefnia et al.\citep{Yousefnia22} applied the Bayesian method to multidimensional experimental observables in high energy heavy ion collisions (HICs), enabling the reconstruction of the impact parameter distributions and the estimation of the variance of observables at a given rapidity or momentum region. 

In Ref.\citep{xiangchen2023}, Chen et al. extended this Bayesian method to low-intermediate energy HICs for reconstructing the impact parameter distribution with two observables, such as the multiplicity of charged particles $M$ and total transverse momentum of light charged particles, $p_t^{tot}$. In this extension, the fluctuation kernel of observable with respect to the impact parameter was assumed as a two-dimensional Gaussian form, and the reconstruction of the impact parameter was achieved. % from two observables, such as the multiplicity of charged particles $M$ and total transverse momentum of light particles, $p_t^{tot}$. 
This advancement allows for a more comprehensive understanding not only on the impact parameter distributions but also on the associated non-Gaussian fluctuation in the observable distribution in low-intermediate energy HICs.

%The Bayesian method which we use is a model-independent method for reconstructing experimental impact parameter distributions, which was proposed by Das~\textit{et al.}\citep{Das18} and further developed in Refs.\citep{Rogly18, Frankland21,Yousefnia22,xiangchen2023}.

%However, when applying the Bayesian method in low-intermediate energy HICs, two issues need to be addressed. One is the influence of the experimental filters on the Bayesian method for reconstructing the impact parameter distributions. The experimental filters are associated with the corresponding setups for detectors, which have physical limitations due to detector placements and designs\citep{tsang2021applying}. These limitations also raise the question of whether the reconstruction of impact parameter distributions with the Bayesian method remains valid again when the experimental filters are implemented. 
More details, the strength of the non-Gaussian fluctuation is measured by the deviations between the measured and reconstructed observable distribution since the reconstructed observable distribution is built from the Gaussian fluctuation kernel. %These deviations quantify the strength of the non-Gaussian fluctuation of the observable distributions.
Theoretically, the deviation can be attributed to the memory effects of the nonequilibrium dynamical process\cite{KaiWen2013} or the large correlation length near the critical point~\cite{BORDERIE201982,Chomaz1999,Chomaz04,YGMa05,AnXin2023,PIETRZAK2020}. As the phenomena associated with the critical point are universal, such as the proliferation of fluctuations, one can also expect that the non-Gaussian fluctuation may appear for nuclear liquid-gas phase transition. %The reason is that the Gaussian nature of the fluctuation is usually related to the system in an equilibrium state. the reconstructed observable distributions were achieved from a series of Gaussian fluctuation kernels based on the Bayes theorem,~\citep{KaiWen2013,AnXin2023}. 

To deeply understand the origins of the fluctuation of observable, i.e., the observable distribution, a comprehensive investigation was conducted within the framework of the improved quantum molecular dynamics model (ImQMD)\citep{Zhang2014,Zhang20FOP} in Ref.\citep{xiangchen2023}. Their results evidence that %the initial fluctuation, mean-field potential and nucleon-nucleon random collisions. 
the initial states of the simulated events have a Gaussian fluctuation nature, which arises from the stochastic sampling of nucleon positions and momenta\citep{Xujun16,Zhang20FOP,HermannPPNP22}. While the nucleonic potential and nucleon-nucleon collisions, which dominate the mechanism of the fragmentation and the mechanism of the liquid-gas phase transition, have the potential to distort the Gaussian fluctuation. Thus, quantitatively analyzing the non-Gaussian fluctuation of the HICs data is essential for us to understand the liquid-gas phase transition in nuclear matter and develop an advanced microscopic transport model. 

In this work, a model-independent method for quantifying the strength of non-Gaussian fluctuations in observables is proposed and its strength is investigated with respect to impact parameter $b$ and beam energy $E_{beam}$ under two different effective nucleon-nucleon interactions. %, which will be helpful in developing the transport model containing the fluctuation term.%the reconstructed observable distribution at a given impact parameter, or named as the fluctuation kernel $P(\mathbf{X}|b)$, has been assumed a Gaussian form. This Gaussian form implies that the  key assumption in 
The paper is organized as follows. Firstly, we generate the `measured' observable distributions with the ImQMD model and filter them with the experimental conditions. Then, the reconstruction of the impact parameter distribution and observable distributions via the Bayesian method are investigated under the experimental conditions.  %The `measured' data of observable distributions is generated from the simulations of ImQMD model, since it contains the impact parameter distribution and can help us understand the validation of this method. 
By comparing the reconstructed observable distributions with `measured' observable distributions, we quantify the deviation between them by using the mean-square-error ($\mathrm{MSE}$). This MSE value also severs as the indicator of the strength of non-Gaussian fluctuations. Our theoretical results show $\mathrm{MSE}$ increases with the beam energy due to the strong nonequilibrium effect of the mid-peripheral collisions.%By measuring the data of $\mathrm{MSE}$ in the future, one may provide a quantitative description of the dynamics fluctuations which will be helpful for developing an advanced model beyond the mean field approximation.

%Our results show that the inclusion of the experimental filters will not change our conclusion, and it verifies the robustness of the Bayesian method in reconstructing the impact parameter distributions. %, we give $\theta$ a range, where $\theta$ is the Angle with the beam direction, stipulate that only fragments in this range can be received. Then we can adopt the Bayesian method to reconstruct the impact parameter distributions from $M$ and $p_t^{tot}$.

The `measured' observables in this study are the multiplicity of charged particles, i.e., $M$, and the total transverse momentum of light charged particles, i.e., $p^{tot}_{t}$, for reaction system $^{112}$Sn+$^{112}$Sn at the beam energies ranging from 50 MeV/u to 200 MeV/u. These data are generated using the ImQMD model(the ImQMD-Sky version)\citep{Zhang2014}. The impact parameter $b$ is randomly distributed within the range from 0 to $b_{max}=1.2 (A_p^{1/3}+A_t^{1/3})$ fm according to the probability density $2b/b_{max}^2$, and the total events number is 1,000,000. $A_p$ and $A_t$ are the mass numbers of projectile and target, respectively. The effective Skyrme interaction SkM* and SLy4 are used in the calculations. More details of this model can be found in Refs.\citep{Zhang2014,Zhang20FOP,HermannPPNP22}.

\begin{figure}[htbp]
	\centering
	\includegraphics[angle=0,scale=0.35,angle=0]{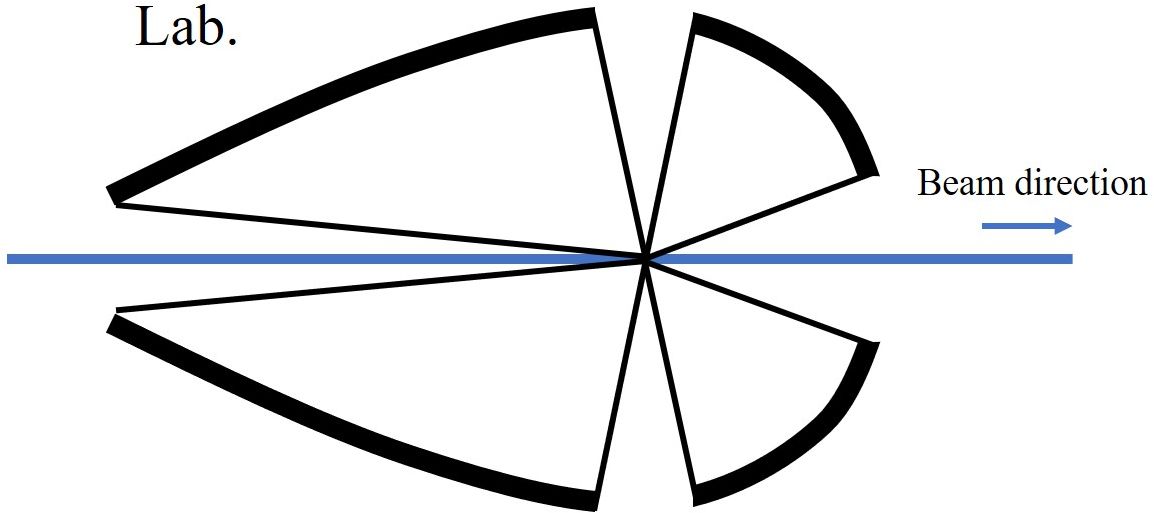}
	\setlength{\abovecaptionskip}{0pt}
	%\vspace{5em}
	\caption{Schematic representation of the experimental detector setup.}
	\label{fig:fig1}
	\setlength{\belowcaptionskip}{0pt}
\end{figure}

To elucidate the impacts of the experimental filters, we use the angle and energy cuts as in Ref.\citep{SARANTITES1996418}, i.e.,  $\theta_{lab}$ from $14^{\circ}$ to $80^{\circ}$ and $100^{\circ}$ to $147^{\circ}$, where $\theta_{lab}$ is the angle in laboratory reference frame, %The geometric coverage is 73.1$\%$ of 4$\pi$, and a sketch plot of the detectors is presented in Fig.~\ref{fig:fig1}. 
and the emitted particles with kinetic energy ranging from 10 MeV to 100 MeV are detected. A sketch plot of the detectors is presented in Fig.~\ref{fig:fig1}. %, and it is named energy-cut, i.e., `E-cut'. 
The reason why we use the transport model to generate the `measured' data is that we know the impact parameter $b$, $p^{tot}_t$, and $M$ for each event. This allows us to check the effectiveness of the Bayesian method on the reconstruction of impact parameter distributions and the form observable distributions at different $b$.

%The `measured' data $P(\mathbf{X})$ used in this work is generated by the transport model, . 
%In addition, in the heavy ion collision experiment, the detector can measure the range of energy is also limited. Therefore, on the basis of $\theta$-cut, we also add an energy cut. that is, limit the energy range that can be measured from 10 MeV to 100 MeV.

%Different geometric coverings give different results. In order for the data used for fitting to be closer to the experimental data, the Angle of the fragments that can be received during the simulation is limited according to the Angle that can be detected experimentally. In this work, We took a range of $\theta_{lab}$ from $14^{\circ}$ to $80^{\circ}$ and $100^{\circ}$ to $147^{\circ}$, where $\theta_{lab}$ is the Angle between the exit direction of the fragment and the beam direction. The geometric coverage is 73.1$\%$ of 4$\pi$. We'll call it $\theta$-cut. Besides, 

%The ImQMD model with $\theta$-cut was used to calculate $^{112}$Sn+$^{112}$Sn at $E_{beam}$=50 MeV/u and 120 MeV/u, and the pseudo-experimental data were generated. The pseudo-experimental data contains the real impact parameter information in the simulation, which can be used to verify the correctness of the Bayesian method. The number of events is 1,000,000, and the impact parameter $b$ is randomly distributed in the range from 0 to $b_{max}=1.2 (A_p^{1/3}+A_t^{1/3})$ fm according to the probability density $2b/b_{max}^2$.

%\section{Brief introduction on the Bayesian method}

%First, we need to obtain pseudo-experimental-data for fitting, that is, the probability density function of the observable vector $X$. In this paper, we use data generated from the ImQMD model\citep{Zhang2014}.

Now, let us first introduce the Bayesian method for model-independently reconstructing impact parameter distributions\cite{Frankland21,Rogly18,Yousefnia22,xiangchen2023}. This method is based on Bayes's theorem,
%The third method which we use involves reconstructing the impact parameter distribution by considering the fluctuations in observables for $b$. 
\begin{equation}
\label{eq:Bayes1}
P(b|\mathbf{X})=\frac{P(b)P(\mathbf{X}|b)}{P(\mathbf{X})}.
\end{equation}
Here, $P(\mathbf{X})$ represents the probability density of the observable $\mathbf{X}$ over the whole impact parameter, which can be measured in experiments. The $P(\mathbf{X}|b)$ denotes the conditional probability density distribution of $\mathbf{X}=(M, p^{tot}_t)$ at a given impact parameter $b$, and is also named as the fluctuation kernel. %Usually, the $P(\mathbf{X}|b)$ is assumed to follow a multidimensional Gaussian form if one assumes the system reaches statistical equilibrium and dynamical process does not dramatically distort the shape of fluctuation\citep{Yousefnia22,xiangchen2023}. 
By utilizing Eq.(\ref{eq:Bayes1}), one can infer the impact parameter distribution at specific values of $\mathbf{X}$, i.e., $P(b|\mathbf{X})$, if the probability density functions(PDFs) $P(b)$ and $P(\mathbf{X}|b)$ are known. 

%However, directly using the Eq.(\ref{Bayes1}) encounters a problem caused by the uncertainties in the overall impact parameter distribution of $P(b)$. 

%In the calculations of $P(b|\mathbf{X})$ according to Eq.(~\ref{eq:Bayes1}), $P(\mathbf{X})$ and $P(\mathbf{X}|b)$ are treated as follows. 
To avoid the uncertainties from the $P(b)$, the $b$-centrality ($c_b$) is introduced as the same as in~\citep{Frankland21}, 
\begin{equation}
\label{eq:cb}
    c_b=\int_0^b P(b') db'.
\end{equation}
This allows for the replacement of the variable $b$, resulting in $P(c_b) = 1$. Thus, a simple form, i.e., 
\begin{equation}
\label{eq:Pxcb}
P(\mathbf{X})=\int P(\mathbf{X}|c_b) dc_b, 
\end{equation}
can be obtained. By fitting the data of $P(\mathbf{X})$, one can obtain the solution for $P(\mathbf{X}|c_b)$ if the form of $P(\mathbf{X}|c_b)$ is given\footnote{which is equal of $P(\mathbf{X}|b)$}. %\footnote{\begin{equation}
% \begin{aligned}
% \label{eq-a-z}
% 		P(c_b|\mathbf{X})P(\mathbf{X}) &= P(\mathbf{X}|c_b)P(c_b)=P(\mathbf{X}|c_b) \\
% 		P(b|\mathbf{X})P(\mathbf{X}) &= P(\mathbf{X}|b)P(b) \\
% 		c_b &= \frac{b^2}{b^2_{max}} \\
% 		\frac{dc_b}{db} &= \frac{2b}{b^2_{max}} \\
% 		P(b|\mathbf{X}) &= P(c_b|\mathbf{X})\frac{dc_b}{db}=P(c_b|\mathbf{X})\frac{2b}{b^2_{max}} \\
% 		P(\mathbf{X}|b) &= \frac{P(b|\mathbf{X})P(\mathbf{X})}{P(b)} \\
% 		&= \frac{P(c_b|\mathbf{X})\frac{2b}{b^2_{max}}P(\mathbf{X})}{\frac{2b}{b^2_{max}}} \\
% 		&= P(\mathbf{X}|c_b)
% \end{aligned}
% \end{equation}}
%, and subsequently, $ P(c_b|\mathbf{X})$ can be obtained using Bayes's theorem. This approach addresses the uncertainties in the impact parameter distribution and allows for a more accurate reconstruction of the impact parameter distribution based on the observed data. %Finally, the expected impact parameter distributions of selected events can be retrieved from $P(b|\mathbf{X}) = P(b)P(c_b|\mathbf{X})$\citep{Frankland21}. 

In this study, the form of $P(\mathbf{X}|c_b)$ is assumed as a Gaussian distribution as same as in Refs.\citep{Yousefnia22,xiangchen2023} and we renamed it as,
\begin{equation}
\label{eq:GS-PDF-cb}
P_G(\mathbf{X}|c_b)=\frac{\exp\{-\frac{1}{2}(\mathbf{X}-\overline{\mathbf{X}}(c_b))^T\Sigma^{-1}(c_b)(\mathbf{X}-\overline{\mathbf{X}}(c_b))\}}{2\pi\sqrt{|\Sigma(c_b)|}}.
\end{equation}
The mean values $\overline{\mathbf{X}}$ and the elements of the covariance matrix $\Sigma_{ij}$ are smooth positive functions of $c_b$, and are expressed as the exponential of a polynomial as in Ref. \citep{Yousefnia22},
\begin{equation}
\begin{aligned}
        \label{eq:mean-var}
        \overline{X}_i(c_b)&=\overline{X}_i(0)\exp\left (-\sum_{n=1}^{n_{\mathrm{max}}} a_{i,n} c_b^n   \right ),\quad 
        i=1, \cdots , N, \\
	\Sigma_{ij}(c_b)&=\Sigma_{ij}(0)\exp\left (-\sum_{m=1}^{m_{\mathrm{max}}} A_{ij,m} c_b^m\right ), \quad
         i=1, \cdots , N.
\end{aligned}
\end{equation}
We rename the reconstructed observable distributions as $Q(\mathbf{X})$, which is, 
\begin{equation}
    Q(\mathbf{X})=\int P_G(\mathbf{X}|c_b) dc_b.
\end{equation}
The values of $n_{max}$, $m_{max}$, $\overline{X}_i(0)$, $a_{i,n}$, $\Sigma_{ij}(0)$ and $A_{ij,m}$ of $P_G(\mathbf{X}|c_b)$ are determined by describing the `measured' data of $P(\mathbf{X})$ with a minimum of the reduced chi-square, i.e., $\chi^2_r\ge \chi^2_{r,min}$. More details, %To use the Bayesian method according to Eq.(\ref{eq:Pxcb}), one has to normalize the distribution of observbales $\mathbf{X}$ to its probability distribution functions (PDFs), i.e., $P(\mathbf{X})$. In addition, 
% \textcolor{blue}{the values of $p^{tot}_{t}$ is reconstructed by the formula $p_{t0}^{tot}=p^{tot}_{t}/(1 \text{GeV/}c)$, for constructing the $\mathbf{X}=\{M, p_{t0}^{tot}\}$ as a dimensionless vector in the Bayesian analysis of multiobservables.}
the values of $M$ and $p^{tot}_{t}$ are rescaled by their maximum values to make their values less than 1, i.e., $M_0=M/M_{max}$ and $p_{t0}^{tot}=p_{t}^{tot}/p_{t,max}^{tot}$, for constructing the $\mathbf{X}=\{M_0, p_{t0}^{tot}\}$ as a dimensionless vector in the Bayesian analysis of multiobservables. The values of $M_{max}$ and $p^{tot}_{t,max}$ obtained with SkM* and different beam energies are listed in the second and third columns of Table~\ref{tab:max}. The normalization procedure does not influence our conclusions as evidenced in Ref.\citep{xiangchen2023}.

\begin{table}[htbp]
\centering
	\caption {The maximum multiplicity of charged particles and the total transverse momentum of light charged particles under the experimental filter for the $^{112}$Sn+$^{112}$Sn at 50 MeV/u and 120 MeV/u. $E_{beam}$ is in MeV/u, and $p_{t,max}^{tot}$ is in GeV/$c$. The last six columns are reduced $\chi^2_{r,n_{max}m_{max}}$ obtained by reconstruction method with different combinations of $n_{max}$ and $m_{max}$.}
 	% \caption {\textcolor{blue}{The reduced $\chi^2_{r,n_{max}m_{max}}$ obtained by reconstruction method with different combinations of $n_{max}$ and $m_{max}$ for the $^{112}$Sn+$^{112}$Sn at 50 MeV/u and 120 MeV/u. $E_{beam}$ is in MeV/u.}}
 
	\begin{tabular}{ccccccccc}
		
		% \hline
		% System  &$E_{beam}$ (MeV/u)  &$M_{max}$   &$p_{t,max}^{tot}$ (GeV/$c$)\\
		% \hline
       
  %             $^{112}$Sn+$^{112}$Sn  &50  &45  &13\\
  %             (with $\theta$ \& E-cut)   &120  &62  &24\\

         \hline
		\hline
		% $E_{beam}$&$M_{max}$&$p_{t,max}^{tot}$ & $\chi^2_{r,11}$ & $\chi^2_{r,22}$ & $\chi^2_{r,23}$ & $\chi^2_{r,32}$ & $\chi^2_{r,33}$ & $\chi^2_{r,44}$\\
  \makebox[0.02\textwidth][c]{$E_{beam}$} 
& \makebox[0.02\textwidth][c]{$M_{max}$}
& \makebox[0.03\textwidth][c]{$p_{t,max}^{tot}$} 
& \makebox[0.01\textwidth][c]{$\chi^2_{r,11}$}
& \makebox[0.01\textwidth][c]{$\chi^2_{r,22}$} 
& \makebox[0.01\textwidth][c]{$\chi^2_{r,23}$} 
& \makebox[0.01\textwidth][c]{$\chi^2_{r,32}$}
& \makebox[0.01\textwidth][c]{$\chi^2_{r,33}$}
& \makebox[0.01\textwidth][c]{$\chi^2_{r,44}$}\\
                % $E_{beam}$ & $\chi^2_{r,11}$ & $\chi^2_{r,22}$ & $\chi^2_{r,23}$ & $\chi^2_{r,32}$
                % & $\chi^2_{r,33}$ & $\chi^2_{r,44}$ \\
		\hline
                
              50  & 45  & 13 & 1.53  &1.21  &1.16  &1.13  &1.10  &1.06 \\
              120  & 62  & 24 & 8.49 &2.09  &1.60  &2.20  &2.15  &4.55\\
                % 50   & 1.53  &1.21  &1.16  &1.13  &1.10  &1.06 \\
                % 120  & 8.49 &2.09  &1.60  &2.20  &2.15  &4.55\\
              
        \hline
        \hline        
		
	\end{tabular}
	\label{tab:max}
\end{table}

\begin{figure}[htbp]
	\centering
	\includegraphics[angle=0,scale=0.4,angle=0,width=1\linewidth]{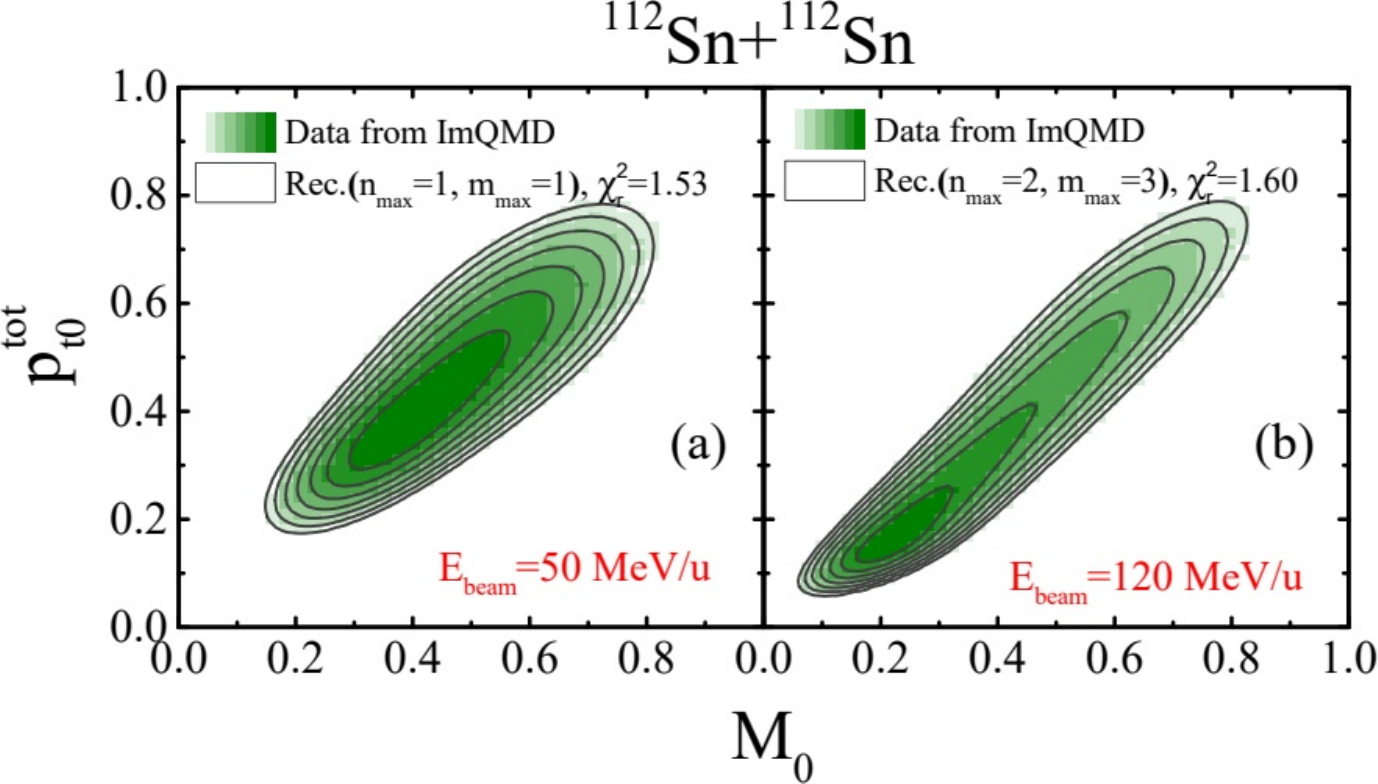}
	\setlength{\abovecaptionskip}{0pt}
	%\vspace{5em}
	\caption{Contour plots of probability density distribution $M_0$ vs $p_{t0}^{tot}$ for $^{112}$Sn+$^{112}$Sn at $E_{beam}$=50 MeV/u (a) and 120 MeV/u (b). The lines are the reconstructed results.}
	\label{fig:fig2-cont-ran}
	\setlength{\belowcaptionskip}{0pt}
\end{figure}

The color maps in Figure.~\ref{fig:fig2-cont-ran} present the $P(\mathbf{X})$s for $^{112}$Sn+$^{112}$Sn, which are generated directly from the ImQMD with SkM* parameter set and serve as the `measured' data. Panel (a) and (b) are the results for the beam energy at 50 MeV/u and 120 MeV/u, respectively. At 120 MeV/u, the largest value of $P(\mathbf{X})$ appears at small $M_0$ 
% \textcolor{blue}{$M$} 
and $p^{tot}_{t0}$ region which corresponds to peripheral collision events. With the beam energy decreasing to 50 MeV/u, the reaction system enters into the spinodal region, where the tiny fluctuation of the system states will dramatically change the final fragmentation pattern. These different fragmentation patterns lead to the strong fluctuation of observables relative to the $b$, and thus the $P(\mathbf{X})$ have a flat distribution over a wide range of $M_0$ 
and $p_{t0}^{tot}$. 

The contour lines in Fig.~\ref{fig:fig2-cont-ran} are the reconstructed results, i.e., $Q(\mathbf{X})$, obtained with the smallest number of $n_{max}$ and $m_{max}$ combinations and smallest $\chi^2_r$ value among the different $n_{max}$ and $m_{max}$ combinations. This reconstruction depends on the solution of the $P_G(\mathbf{X}|c_b)$, which requires to determine the 6 groups of parameters in Eq.(\ref{eq:mean-var}), such as $\overline{X}_i(0)$, $a_{i,n}$, $\Sigma_{ij}(0)$, $A_{ij,m}$, $n_{\mathrm{max}}$, and $m_{\mathrm{max}}$. The index $i$ ranges from 1 to $N$, $n$ from 1 to $n_{max}$, and $m$ from 1 to $m_{max}$. Thus, the total number of parameters $N_{para}$ is,
\begin{equation}
    N_{para}=N(n_{max}+1)+\frac{N(N+1)}{2}(m_{max}+1).
\end{equation}
In this work, $N=2$. $n_{max}$ and $m_{max}$ can take different combinations. When $n_{max}=1$ and $m_{max}=1$, $N_{para}=10$ and when $n_{max}=4$ and $m_{max}=4$, $N_{para}=25$. These parameters are determined by fitting the `measured' data $P(\mathbf{X})$ with the MINUIT\citep{James1994MINUITFM}. %For the convenience of description, we named the reconstructed observables distributions as $Q(\mathbf{X})$. %The lines in Fig.~\ref{fig:fig2-cont-ran} are the fitting results obtained with the conditions that $N_{para}$ take the smallest values at $\chi^2_{r}<2$. %These values are smaller than 3 (\textcolor{red}{$n_{max}$=4,$m_{max}$=4, E=120MeV/u, $\theta-cut$,$\chi^2_r$=4.55}), while the values of $\chi^2_r$ in Ref.~\citep{Yousefnia22} are greater than 2.9. It means that the Gaussian fluctuation kernel is a better approximation for low-intermediate energy HICs than high energy HICs.

% \begin{table}[htbp]
% 	\caption {$\chi^2_r$ obtained by reconstruction method in the $^{112}$Sn+$^{112}$Sn system with  $\theta$ \& E-cut at different beam energies and different combinations of $n_{max}$ and $m_{max}$.}
% 	\begin{tabular}{c|ccccccc|ccccccccccc}

% 		\hline
% %	$E_{beam}$ (MeV/u)    &\multicolumn{7}{c}{$\chi^2_r$} \\
% 		\hline
% 		\diagbox[]{Filters}{$\chi^2_r$}{Para.} & (1,1)  & (2,2)  & (2,3)  & (3,2)  & (3,3)  & (4,4)  & & $E_{beam}$   \\  
%         \hline
% 		$\theta$ \& E-cut    &1.53  &1.21  &1.16  &1.13  &1.10  &1.06  & & 50 MeV/u\\
%         \hline
%             $\theta$ \& E-cut &8.49 &2.09  &1.61  &2.20  &2.15  &4.55 & & 120 MeV/u\\   
%         \hline
%         \hline
% 	\end{tabular}
% 	\label{tab:chi2s}
% \end{table}

However, one should keep in mind that it is hard to justify how many fitting parameters are good enough by only seeking the minimum of $\chi^2_r$ among the different parameter sets, since the real $b$ dependence of $\overline{X}_i$ and $\Sigma_{ij}$ or the real $b$ distribution are not known in advance in experiments. Thus, we also list the values of $\chi^2_r$ obtained with other combinations of $n_{max}$ and $m_{max}$ in Table~\ref{tab:max} which will be used to estimate the uncertainties of the reconstructed method. 

\begin{figure}[htbp]
	\centering
	\includegraphics[angle=0,scale=0.4,angle=0,width=1\linewidth]{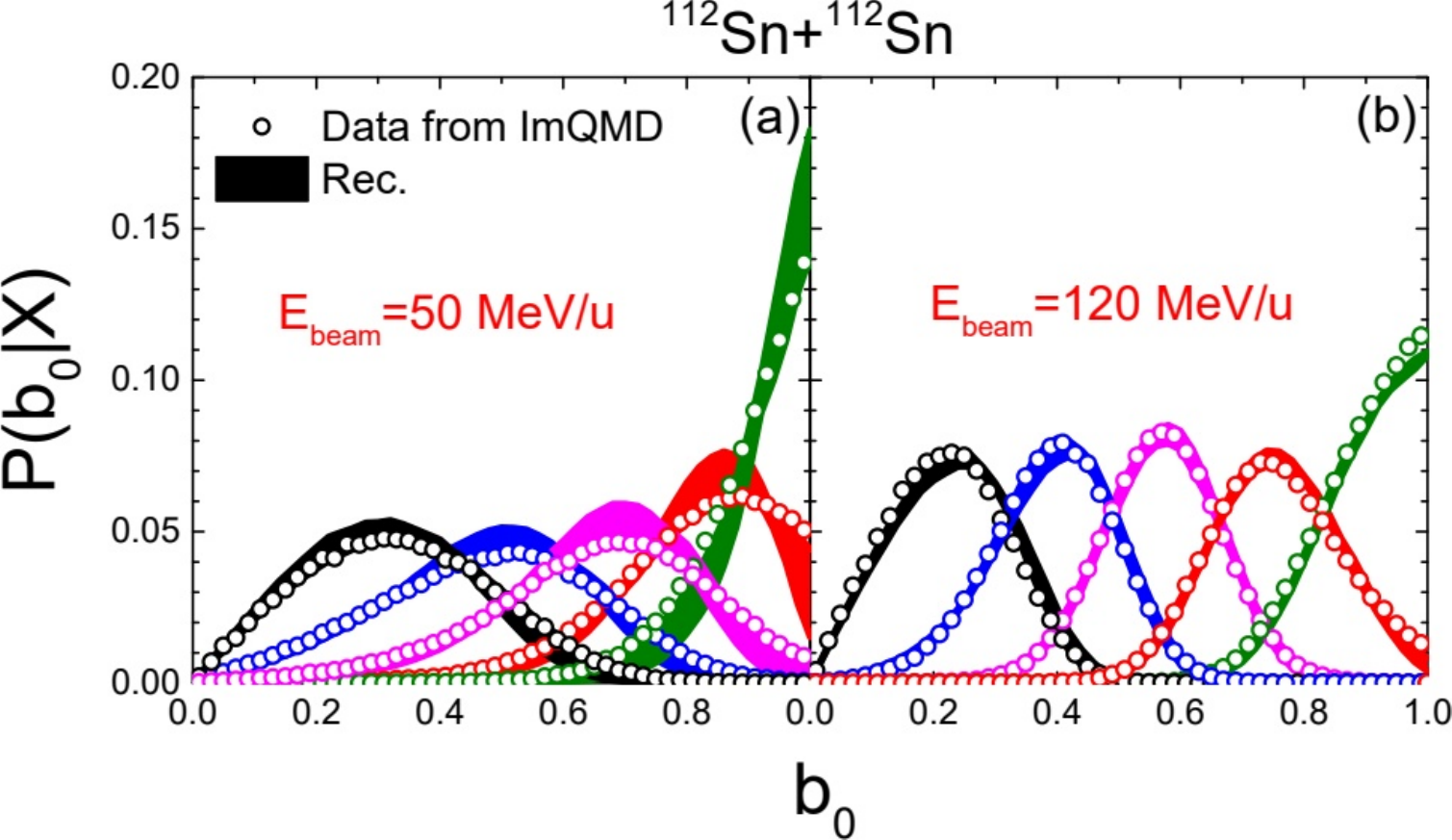}
	\setlength{\abovecaptionskip}{0pt}
	%\vspace{5em}
	\caption{Reduced impact parameter distributions for five clusters; open circles are the impact parameter distributions obtained from ImQMD model with SkM*, and shaded regions are the results inferred with the reconstructing method with different $n_{max}$ and $m_{max}$ combinations. }
	\label{fig:fig6}
	\setlength{\belowcaptionskip}{0pt}
\end{figure}

In Fig.\ref{fig:fig6}, we plot the predicted reduced impact parameter $b_0=b/b_{max}$ distributions (shaded regions), i.e., $P(b_0|\mathbf{X})$, by using the Bayesian method, i.e.,
\begin{equation}
    \label{eq:recon}
    \begin{aligned}
        P(b|\mathbf{X} \in C_k)&=P(b)P(c_b|\mathbf{X} \in C_k)\\
        &=\frac{P(b)\int_{\mathbf{X}\in \Omega(C_k)} P_G(\mathbf{X}|c_b)d\mathbf{X}}{\int_{\mathbf{X}\in \Omega(C_k)} P(\mathbf{X}) d\mathbf{X}}. %=\frac{P(b)}{\Delta C_k} \int_{\mathbf{X} \in C_k} P(\mathbf{X}|b) d\mathbf{X}
    \end{aligned}   
\end{equation}
%$P(\mathbf{X}|b)$ comes from the reconstruction method of different combinations of $n_{max}$ and $m_{max}$ (shaded areas) by using \textcolor{red}{$P(\mathbf{X}|b)\approx P(\mathbf{X}|c_b)b^2/b_{max}^2$}. 
$C_k$ is the subdataset of the $i$th classification of data $\mathbf{X}$ by using the $K$-means method as same as in Ref.\citep{xiangchen2023}. 
The shaded bands in the figure are the results obtained with $\chi_r^2<\chi_{r,min}^2+1$ for different combination of  $n_{\mathrm{max}}$ and $m_{\mathrm{max}}$. As detailed in Table~\ref{tab:max}, all the values of $\chi^2_{r,n_{max}m_{max}}$ obtained with our selected combinations are less than $\chi_{r,min}^2+1$ for $E_{beam}$=50 MeV/u. For $E_{beam}=120$ MeV/u, there are four combinations of $n_{max}$ and $m_{max}$ that yield the $\chi^2_{r,n_{max}m_{max}}<\chi_{r,min}^2+1$. It indicates larger uncertainties of $n_{max}$ and $m_{max}$ for $E_{beam}=50$ MeV/u than that for $E_{beam}=120$ MeV/u,
%At beam energy of 50 MeV/u, the $P(\mathbf{X})$ distributed in a more flatten shape than that in 120 MeV/u, which has been shown in Fig.\ref{fig:fig2-cont-ran}. 
%Thus, the number of combination $n_{\mathrm{max}}$ and $m_{\mathrm{max}}$ for fitting $P(\mathbf{X})$ at 50 MeV/u is larger than that with 120 MeV/u, 
and result in a wider uncertainty region of the reconstructed impact parameter distributions as in panel (a) than that in panel (b). Nonetheless, the figure illustrated that the predicted reduced impact parameter distributions from the reconstructing method agree well with the actual impact parameter distributions (open circles) under the different combinations of $n_{\mathrm{max}}$ and $m_{\mathrm{max}}$. %These reconstructed impact parameter distributions will be helpful for us to understand the strength of the non-Gaussian fluctuation as a function of impact parameter or centrality in experiments. 
The high fidelity of the reconstructed impact parameter distributions allows us to understand the strength of the non-Gaussian fluctuation as a function of impact parameter or centrality in experiments.

%With the K-means algorithm, we can classify events by $\mathbf{X}=\{M_0, p_{t0}^{tot}\}$\citep{xiangchen2023}. The events points are sorted into five clusters by the K-means clustering algorithm. 

%The reasons why the reconstructing method can reproduce the real impact parameter distribution and is less influenced by the deviations of the covariance matrix can be understood from the following two aspects: one is the validity of Gaussian assumptions on the $P(\mathbf{X}|b)$, and another is the reconstructing method on $P(b|\mathbf{X})$ based on Eq. (\ref{eq:recon}).

%\subsection{Impact parameter dependence of the non-Gaussian fluctuation of $M$ and $p_{t}^{tot}$}
%\label{sec:fluctuation}

%Now, let's turn to quantify the deviation between the reconstructed observable distributions $Q(\mathbf{X})$ and measured distributions $P(\mathbf{X})$ at different impact parameters $b$. 
In general, the fluctuations of observables with respect to $b$ stem from the addition of the random sources of the initialization and nucleon-nucleon scattering on smaller scales\citep{xiangchen2023}. The assumption on the Gaussian form of $P(\mathbf{X}|c_b)$ implies that the systems reach equilibrium since the observables from different events and different impact parameters are independent of each other. %Thus, the reconstructed observable distributions based on the Gaussian fluctuation kernel, i.e., $P_G(\mathbf{X}|c_b)$, could be different than the real distribution $P(\mathbf{X})$. %To distinguish the reconstructed observable distributions from the `measured' observable distributions $P(\mathbf{X})$, we renamed the reconstructed observable distributions as $Q(\mathbf{X})$, which is obtained from $P_G(\mathbf{X}|c_b)$, 
% \begin{equation}
%     Q(\mathbf{X})=\int P_G(\mathbf{X}|c_b) dc_b.
% \end{equation}
Thus, the deviation between $P(\mathbf{X})$ and $Q(\mathbf{X})$ quantifies the strength of non-Gaussian fluctuation, and can reflect the nonequilibrium effect of heavy ion collisions. Nonetheless, the obtained $\chi^2_r$ can not completely measure the deviation between $P(\mathbf{X})$ and $Q(\mathbf{X})$. The reason is that the $\chi^2_r$ is obtained only within the `measured' data region, i.e., the region with $P(\mathbf{X})\ne 0$, and does not consider the difference of two distributions in the region where $P(\mathbf{X})=0$ but $Q(\mathbf{X})\ne 0$.

%and are often approximated by the Gaussian distributions. The deviation may be caused by the non-equilibrium collision process. Thus, it will be interesting to discuss the strength of the non-Gaussian fluctuation of $\mathbf{X}=\{M_0, p_{t0}^{tot}\}$ for low-intermediate energy HICs. %which will be quantitatively described by the mean-square-error (MSE). %Kullback-Leibler divergence ($D_{KL}$)\citep{Goodfellow-et-al-2016}.  %the variance and covariance of the distribution $P(\mathbf{X})$, at different impact parameters. 

%Quantitatively, the goodness of the reconstructed impact parameter distribution can be described by 
%To avoid the zero probability issue for the reconstructed observable distributions, we propose a modified Kullback-Leibler divergence ($D^*_{KL}$) in this work, which is defined as

Usually, the cumulants of the observable are used to quantify the non-Gaussian fluctuations in the field of high energy HICs for investigating the phase transition~\citep{Stephanov2009,CZhou2017,Rupam2023}. However, the relation between different cumulants becomes much more complicated for two observables than only using one observable. Thus, we use the mean-square-error (MSE), which is defined as,
% \begin{align}
% \label{eq:DKL}
%     D^*_{KL} =
%         \begin{cases}
% 		D_{KL}(P||Q) & Q(\mathbf{X})\ne 0, \\
% 		D_{KL}(Q||P) & Q(\mathbf{X})=0,
% 	\end{cases}
% %      D^*_{KL}(P||Q)&=\int P(\mathbf{X})\log(f(\mathbf{X}))d\mathbf{X}\\\nonumber
% %       &=\sum_{i=1}^N P(\mathbf{X}_i)\log(f(\mathbf{X}_i))\Delta\mathbf{X}_i. 
% \end{align}
\begin{equation}
\label{eq:mse}
    \mathrm{MSE}(P||Q)=\frac{1}{S}\sum_{i=1}^S \left(P(\mathbf{X}_i)-Q(\mathbf{X}_i)\right)^2,
\end{equation}
%\begin{align}
%	\begin{cases}
%		P(\mathbf{X})log(\frac{P(\mathbf{X})}{Q(\mathbf{X})})\quad & Q(\mathbf{X})\ne 0, \\
%		Q(\mathbf{X})log(\frac{Q(\mathbf{X})}{P(\mathbf{X})}) & Q(\mathbf{X})=0,\\
%	\end{cases}
%\end{align}
to quantify the strength of the non-Gaussian fluctuations. %\textcolor{red}{It is different than the $\chi^2_r$ in the reconstructing procedure, $\mathrm{MSE}$ reflect the different between $P(\mathbf{X})$ and $Q(\mathbf{X})$.} 
$P(\mathbf{X}_i)$ and $Q(\mathbf{X}_i)$ in Eq.(\ref{eq:mse}) represent the `measured' and reconstructed probability density distribution for $\mathbf{X}$ at mesh point $i$ ($i=1, \cdots, S$) in a unitary area, respectively. $S$ is the number of the mesh points in the whole area. The factor $1/S$ is used to normalize the result. %The $P(\mathbf{X}_i)$ and $Q(\mathbf{X}_i)$ are obtained over the whole impact parameter region. 

\begin{figure}[htbp]
	\centering
	\includegraphics[angle=0,scale=0.35,angle=0,width=1\linewidth]{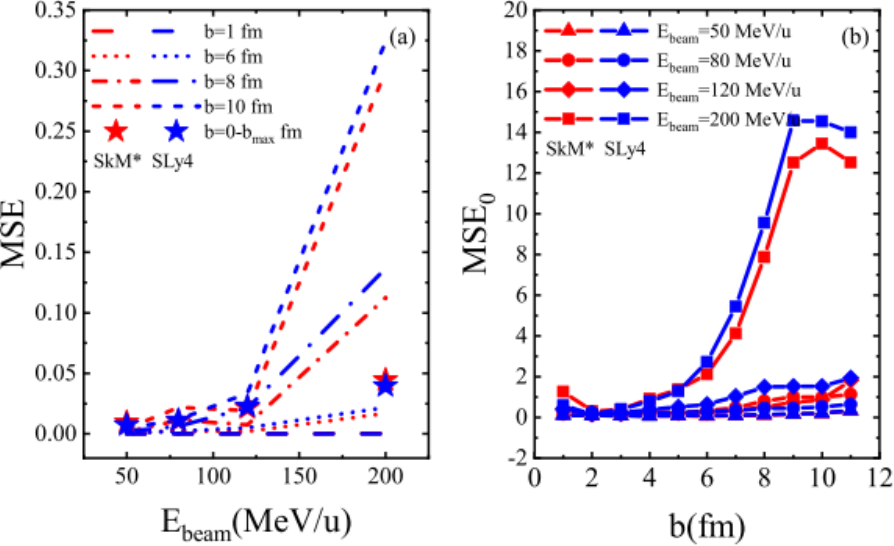}
	\setlength{\abovecaptionskip}{0pt}
	%\vspace{5em}
	\caption{Panel (a) MSE as functions of beam energy. Stars are the results obtained in the whole impact parameter region, and lines are the results for $b$=1, 6, 8, and 10 fm (lines with different styles); (b) $\mathrm{MSE}_0$ as functions of $b$ at different beam energies .}
	\label{fig:fig4-D-KL}
	\setlength{\belowcaptionskip}{0pt}
\end{figure}

Fig.\ref{fig:fig4-D-KL} (a) shows the $\mathrm{MSE}$ %, which is obtained over the whole impact parameter region (green stars), 
as a function of beam energy within the range of 200 MeV per nucleon (red and blue stars). Generally, the values of $\mathrm{MSE}$ increase with beam energy for two kinds of effective interactions, which is caused by the decreasing of the equilibrium degree of the system as the beam energy increases. To deep understand it, we also analyze the $\mathrm{MSE}$ as a function of $E_{beam}$ at $b$=1, 6, 8, and 10 fm, which are calculated by using Eq.(\ref{eq:mse}) but the $P(\mathbf{X}_i)$ and $Q(\mathbf{X}_i)$ are replaced by the conditional probabilities $P(\mathbf{X}_i|b)$ and $P_G(\mathbf{X}_i|b)$ at given $b$.
%which are represented as dashed lines with different colors. In the calculations of $\mathrm{MSE}$ at given $b$, Eq.(\ref{eq:mse}) is also used . 
As illustrated in the figure, the $\mathrm{MSE}$ weakly increases with the $b$ increasing from central collisions to peripheral collisions at the beam energy less than 120 MeV/u. %The reason is that there are many more nucleons colliding each others for central collisions than peripheral collisions, and a higher degree of equilibrium will be reached. Thus, the Gaussian shape of fluctuation of the observable distribution remains intact. 
This increasing trend is caused by the growing anisotropy with impact parameter increasing in the overlapped region, which makes the emitted particles carry more entrance channel information. Thus, their distributions tend to deviate from the Gaussian shape, and the values of MSE become larger than those for central collisions due to the stronger nonequilibrium effect. %which transforms into an anisotropy in momentum space over time due to the pressure gradient between the overlapped and spectator regions. 
With the beam energy increasing above 120 MeV/u, the $b$ dependence of $\mathrm{MSE}$ becomes stronger since the nonequilibrium effect becomes more strong, especially for $b>6$ fm.

However, the impact parameter weight is taken into account in the calculations of $\mathrm{MSE}$ at different $b$ in panel (a), which may lead a misunderstanding on the strength of non-Gaussian fluctuation at different $b$. To single out the strength of non-Gaussian fluctuation from the impact parameter weight at different $b$, a normalized $\mathrm{MSE}_0$ is used, which reads
\begin{equation}
\label{eq:mse0}
    \mathrm{MSE}_0(P||Q,b)=\frac{1}{S}\sum_{i=1}^S \left(\frac{P(\mathbf{X}_i|b)\Delta b}{2 b\Delta b/b^2_{max}}-\frac{Q(\mathbf{X}_i|b)\Delta b}{2 b\Delta b/ b^2_{max}}\right)^2.
\end{equation}
The factor $ 2 b\Delta b/b^2_{max}$ comes from the relationship, 
\begin{equation}
    \int_{b}^{b+\Delta b}\int P(\mathbf{X}|b) d\mathbf{X}db\approx 2\pi b\Delta b/\pi b^2_{max}.
\end{equation}
The $\mathrm{MSE}_0$ as functions of $b$ are plotted in panel (b). 
When $E_{beam}=50,80,120$ MeV/u, the value of $\mathrm{MSE}_0$ monotonously increase with the impact parameter $b$. At $E_{beam}=200$ MeV/u, the values of $\mathrm{MSE}_0$ reach the maximum at b=9 fm and then decreases with impact parameter increase for SkM*%SLy4
. The maximum value of $\mathrm{MSE}$ reflects that the stongest nonequilibrium effects, and it could be related to the pseudocritical point in the liquid-gas phase transition for finite systems. Because the long correlation length near the critical point will lead to a non-Gaussian shape of observable distribution. To understand the influence of effective Skyrme interaction on the position of the maximum $\mathrm{MSE}_0$, we also calculate the $\mathrm{MSE}_0$ as a function of $b$ with the parameter set SLy4. Our calculations show that $\mathrm{MSE}_0$ reaches a maximum at b=9 fm, which is smaller than that obtained with SkM*. But the difference of $\mathrm{MSE}_0$ obtained with SkM* and SLy4 is small. It implies that the critical temperature for SLy4 and SkM* is different and the difference is small. This trend is consistent with the theoretical predictions\cite{RIOS2010}.

% The interesting finding is that the $\mathrm{MSE}_0$ reaches a maximum around $b=5$ fm, and then decreases with impact parameter increase. These results evidence that the nonequilibrium effects reach the maximum for mid-peripheral collisions and could be related to the pseudocritical point in the liquid-gas phase transition for finite system. 

% due to the mixing of the reaction mechanism.
%$\Delta \mathbf{X}=dX_1dX_2$ is the area element for calculating the probability in the area region labeled with $\mathbf{X}_i$, which is used to avoid the area element dependence on the calculations of $D_{KL}$. If the reconstructed method can exactly reproduce the measured probability density distributions, $D_{KL}=0$. The non-zero values of $D_{KL}$ reflect the strength of the fluctuation deviating away from the Gaussian distributions. Furthermore, the $D_{KL}$ also reflects the difference of the information entropy between the measured observables distributions and reconstructed Gaussian distribution.%Our calculations show that the values of $D_{KL}$ obtained with different combinations of ($n_{max}$, $m_{max}$)  are less than 0.16.
%In physics, the $D_{KL}$ reflects the difference of the information entropy between the measured observables distributions and reconstructed Gaussian distribution.

% \textcolor{blue}{Furthermore, we calculate the MSE and MSE$_0$ when the effective Skyrme interaction is SLy4, also shown in Fig.\ref{fig:fig4-D-KL}. It can be seen that the results of SLy4 are similar to those of SkM*.}

In summary, the validation of the Bayesian method for reconstructing the impact parameter distribution from the `measured' multiplicity of charged particles and total transverse momentum of light-charged particles under the experimental filters, such as the angle and energy cuts, are tested. Our results show that the experimental conditions do not affect the reliability of the Bayesian method in the reconstruction of impact parameter distribution. 

Further, we investigate the deviation of `measured' observables distribution $P(\mathbf{X})$ away from the reconstructed observables distribution $Q(\mathbf{X})$ via the Bayesian method. % which is obtained over the summation of Gaussian shape fluctuation kernels. 
This deviation reflects the strength of the non-Gaussian fluctuation and can be quantified by the mean-squre-error ($\mathrm{MSE}$). Our calculations show that the $\mathrm{MSE}$ increases with the beam energy, which is mainly caused by the nonequilibrium dynamics for low-intermediate energy heavy ion collisions. At the beam energy of 200 MeV/u, the non-Gaussian fluctuation reach maximum at b=9-10 fm for Sn+Sn, which may be related to the long correlation length and can be used to probe the pesudocritical point for liquid-gas phase transition. It will be exciting to see the experimental results on the $\mathrm{MSE}$, as it can offer valuable insights into nonequilibrium dynamics, and be helpful for estimating the fluctuation caused by high-order correlations and the interplay between geometric symmetry, collective motion, and stochastic nucleon-nucleon collisions in nuclear collisions.  %Our calculations show that the intensity of nonequilibrium dynamic fluctuations increases with the increase of energy. %Finally, we propose this Bayesian method to extract the intensity of non-Gaussian fluctuation or dynamic fluctuations when experimental data are available, which will be of great significance to considering the dynamical fluctuation in the transport models and developing the advanced transport theoretical model in the future.

\section*{Acknowledgments}

This work was partly inspired by the transport model evaluation project, and it was supported by the National Natural
Science Foundation of China under Grants No. 12275359, No. 12375129, No. 11875323 and No. 11961141003, by the National Key R\&D Program of China under Grant No. 2023 YFA1606402, by the Continuous Basic Scientific Research Project, by funding of the China Institute of Atomic Energy under Grant No. YZ222407001301, No. YZ232604001601, and by the Leading Innovation Project of the CNNC under Grants No. LC192209000701 and No. LC202309000201. We acknowledge support by the computing server SCATP in China Institute of Atomic Energy.

% \section{Title 2}
% %%\label{}
% \lipsum[1]

% \subsection{Subsection title}

% \begin{figure}
% 	\centering 
% 	\includegraphics[width=0.4\textwidth, angle=-90]{Physics_Letters_B.pdf}	
% 	\caption{Physics Letters B journal cover} 
% 	\label{fig_mom0}%
% \end{figure}

% A random equation, the Toomre stability criterion:

% \begin{equation}
%     Q = \frac{\sigma_v \times \kappa}{\pi \times G \times \Sigma}
% \end{equation}

% \section{Title 3}
% %%\label{}
% \lipsum[2]

% \subsection{Subsection title}
% \lipsum[3]

% \begin{table}
% \begin{tabular}{l c c c} 
%  \hline
%  Source & RA (J2000) & DEC (J2000) & $V_{\rm sys}$ \\ 
%         & [h,m,s]    & [o,','']    & \kms          \\
%  \hline
%  NGC\,253 & 	00:47:33.120 & -25:17:17.59 & $235 \pm 1$ \\ 
%  M\,82 & 09:55:52.725, & +69:40:45.78 & $269 \pm 2$ 	 \\ 
%  \hline
% \end{tabular}
% \caption{Random table with galaxies coordinates and velocities, Number the tables consecutively in
% accordance with their appearance in the text and place any table notes below the table body. Please avoid using vertical rules and shading in table cells.
% }
% \label{Table1}
% \end{table}

% \section{Discussion}
% %%\label{}
% \lipsum[4]

% \section{Summary and conclusions}
% %%\label{}
% \lipsum[1-4]

% \section*{Acknowledgements}
% Thanks to ...

% %% The Appendices part is started with the command \appendix;
% %% appendix sections are then done as normal sections
% \appendix

% \section{Appendix title 1}
% %% \label{}

% \section{Appendix title 2}
%% \label{}

%% If you have bibdatabase file and want bibtex to generate the
%% bibitems, please use
%%
%\bibliographystyle{elsarticle-num} 
\bibliographystyle{unsrt} 
\bibliography{example}

%% else use the following coding to input the bibitems directly in the
%% TeX file.

%%\begin{thebibliography}{00}

%% \bibitem[Author(year)]{label}
%% For example:

%% \bibitem[Aladro et al.(2015)]{Aladro15} Aladro, R., Martín, S., Riquelme, D., et al. 2015, \aas, 579, A101

%%\end{thebibliography}
\end{CJK}
\end{document}